\documentclass[12pt,preprint]{aastex}

\slugcomment{Revised version}

\usepackage{amsfonts}       
\usepackage{bm}             
\usepackage{color}          
\usepackage{dcolumn}        
\usepackage{graphicx}       
\usepackage{multirow}       
\usepackage{multirow}       
\usepackage{natbib}         
\usepackage{rotating}       

\newcommand{\be}{\begin{equation}}
\newcommand{\cm}{{~\:\rm cm}}

\newcommand{\dtot}[2]{\frac{d{#1}}{d {#2}}}
\newcommand{\ee}{\end{equation}}
\newcommand{\g}{{\bm g}}
\newcommand{\K}{{\rm K}}
\newcommand{\keV}{{~\:\rm keV}}
\newcommand{\km}{{~\:\rm km}}
\newcommand{\kms}{{~\:\rm km\, s^{-1}}}
\newcommand{\kpc}{{~\:\rm kpc}}
\newcommand{\msun}{~\: M_\odot}
\newcommand{\msyr}{{~\:M_\odot\;\rm yr^{-1}}}
\newcommand{\pc}{{~\:\rm pc}}

\renewcommand{\v}{{\bm v}}
\newcommand{\yr}{{~\:\rm yr}}

\begin{document}



\title{On the Nature of Feedback Heating in Cooling Flow Clusters}


\author{Fabio Pizzolato}
\affil{Department of Physics, Technion--Israel Institute of Technology,
Haifa 32000}
\email{fabio@physics.technion.ac.il}

\and

\author{Noam Soker}
\affil{Department of Physics, Technion--Israel Institute of Technology,
Haifa 32000}
\email{soker@physics.technion.ac.il}
 


\begin{abstract}
We study the feedback between heating and cooling of the
intra-cluster medium (ICM) in cooling flow (CF) galaxies and clusters.
We adopt the popular view that the heating is due to an
active galactic nucleus (AGN), i.e.  a central black hole
accreting mass and launching jets and/or winds.
We propose that the feedback occurs with the entire
cool inner region ($r \lesssim 5-30 \kpc$), where the
non-linear over-dense blobs of gas with a density contrast
$\rho/\rho_a \gtrsim 2$  cool fast and are removed
from the ICM before experiencing the next major AGN heating event.
We term this scenario {\it cold-feedback}.
Some of these blobs cool and sink toward the central black hole,
while others might  form stars and cold molecular clouds.

We derive the conditions under which  the dense blobs formed by perturbations
might cool to low temperatures ($T < 10^4~K$), and feed the black hole.
The main conditions are found to be:
(1)~An over-dense blob must be prevented from  reaching an
equilibrium position in the ICM: therefore it has to cool fast,
and the density profile of the ambient  gas should be shallow;
(2)~Non-linear perturbations are required: they might have chiefly
formed by previous AGN activity;
(3)~The cooling time of these non-linear perturbations
should be short relative to few times the typical interval between
successive AGN outbursts.
(4) The blobs should be magnetically disconnected from their
surroundings, in order not to be evaporated by thermal conduction.
\end{abstract}




\section{Introduction}
\label{s:intro}

The observations of Galaxy Clusters with the last generation
X-ray satellites {\sl Chandra} and {\sl XMM-Newton} have shown
the remarkable lack of the large amount of cool gas predicted by
the old version (ten years ago) of the  cooling flow (CF) 
model  \citep[e.g.][]{Pet03, Tam01a, Mol01}.
The most straightforward explanation is that the intra-cluster medium (ICM)
in CF clusters must be  heated by some mechanism
\citep[e.g.][]{Fab03}.
Probably the most popular heating engine is an active galactic nucleus
(AGN) residing at the center of the cluster dominant  galaxy, often a cD
(see e.g. \citeauthor*{Bin95} 1995;  \citeauthor*{Tuc97} 1997;
\citeauthor*{Cio01} 2001; \citeauthor*{Bin04} 2004;
for more references see \citeauthor*{Pet04} 2004 and
the review by \citeauthor*{Beg04} 2004;  see
\citeauthor*{Fuj04} 2004 and \citeauthor*{Fab02} 2002
for some problems connected with this scenario).
Other heating mechanisms, like supernovae and heat conduction
from the cluster outer regions were shown to be problematic
for several reasons.
It is sufficient to mention that we are looking for a unified
mechanism to heat the ICM in CFs, from galaxies to large clusters.
Heat conduction cannot work in galactic CFs because there are no
large heat reservoirs, and probably there are not enough
supernovae in elliptical  galaxies to supply a sufficient amount of heat.

The new X-ray observations show  that the
mass cooling rate to low temperatures is far below predictions
by older versions of the cluster CF model \citep[][]{Fab94},
but is compatible with low-mass cooling rate models
\citep[see][on the expectation for that result]{Bin04}.
Among them, the moderate CF model
\citep[][]{Sok01, Sok03, Sok04a}
is different from many earlier proposed
processes whose aim is to prevent CF in clusters of galaxies
altogether. In the moderate CF model small but non-negligible quantities
of gas are cooling to low ($\lesssim 10^4$~K) temperatures.

Most models of AGN heating agree in that there is some sort of
feedback between the heating and the radiative cooling.
Many of these models result in intermittent AGN activity.
There are two approaches to the feedback between the cooling
ICM and  the AGN. In the first, the ICM does not cool
below X-ray emitting temperatures;  the AGN
accretion is Bondi-like, and is determined by the ICM properties
{\it very} close to the central
black hole \citep[e.g.][]{Chu02, Nul04, Omm04}.
\citet{Nul04} for example, studies the accretion of gas at the virial
temperature. We term this type of models {\it hot feedback}.

In the second approach, the black hole accretes cold gas, but
the mass cooling rate is much below that in
old versions of the cooling flow model.
In this case the feedback takes place within  a region
extending to a distance of $\approx 5-30\kpc$ from the cluster center.
Such is the moderate cooling flow model
\citep[][]{Sok01, Sok04a}.
We term this {\it  cold feedback} model.
The two types of feedback models imply other differences.

\begin{enumerate}

\item
In the hot feedback models the optical filaments
observed in many CF-clusters \citep[][]{Hec89},
and the cooler molecular gas detected via CO observations
(\citeauthor{Edg01} 2001; also \citeauthor{Edg03} 2003;
\citeauthor{Sal03} 2003; \citeauthor{Sal04} 2004)
come from stripping gas from cluster galaxies.
In the cold feedback models the cold gas may come from
the cooling ICM as well.

\item
In the cold feedback models  some X-ray emission from gas at
temperatures $\lesssim 10^7$~K is predicted, but at a level more
than an order of magnitude below that in old versions of
the cooling flow model
\citetext{Pizzolato, Behar, \& Soker in preparation}.

\item
The feeding of the central black hole with cold gas in the
cold feedback models  makes the process similar in some
aspects to that of AGN in spiral galaxies.

\item
It is possible, although not required, that in the cold feedback model
the feedback does not only keep energy balance, but mass balance as well.
Namely, a substantial fraction of gas that cooled to low temperature is
injected back to the ICM and heats up as it is shocked \citep[][]{Sok05}.
It is possible that during a fraction of the time, most
of the cooling gas forms molecular clouds and stars, rather than
being injected back to the ICM at high speeds.
\end{enumerate}

To set the stage for the discussion to follows, it will be constructive
to present the temperature profiles of some CF clusters 
(Figure~\ref{f:tprof}).
In the presently proposed model a flat temperature profile in the inner
region is expected, at least during some fraction of the duty cycle.
It is that flat and low temperature profile which facilitates the formation
of cold blobs.
This is contrary to some hot feedback models, where  the accreted gas
comes from the immediate AGN neighborhoods.
It is our opinion that some of these models may have problems in
accounting for the temperature profiles presented in Figure~\ref{f:tprof}.
For example,  the repeated heating by \citet{Omm04}
does not heat the outer region as much as  the inner one.
It seems as if the outer region will cool to low temperature,
unlike the higher temperature of the outer regions in clusters.
\citet{Rus02} take both AGN heating and heat conduction.
We find their temperature profile in the inner several~kpc
region to be too steep (see their Figure~1) compared with  real clusters.
In the temperature profile obtained by \citet[their Figure~5]{Hoe04} 
we see two problems.
First, contrary to their claim, we do not think that in the outer CF
region
their profile fits that of A2052 \citep[taken from][]{Bla03}.
Second, after a time of $\approx 1.2 \times 10^{10}$ years, the
temperature in the inner $\approx 2 \kpc $ drops to $\ll 10^7~\K$,
contrary to the observations.
There is a fine tuning problem, in that presently all clusters
do not show this drop, but they will in a few $10^9$~years.

The considerations above, among other arguments,
motivate us to consider the cold feedback model.
In the present paper we examine the two types of feedback heating.
By analyzing published models and by comparing them with the observations,
we constrain the parameters space of the different models, and make some
predictions which can be tested with future observations.
Our proposed scenario is presented in \S~\ref{s:scenario}.
In \S~\ref{s:nnlin}  we calculate the evolution of a dense blob,
assuming the dense blobs are magnetically disconnected from
their environment, e.g., as in the magnetic flux loop
model of \cite{Sok04a} where heat conduction occurs only within
the cold blob.
Readers interested only in the basic scenario, results, and predictions,
can skip \S~\ref{s:nnlin}  and go directly from \S~\ref{s:scenario} to our 
summary in \S~\ref{s:summary}.

\section{The Proposed Cold Feedback Scenario}
\label{s:scenario}

The cold feedback scenario entails a cycle in the cooling/accretion
activity. We suggest that this cycle starts with a major AGN outburst,
which injects a huge amount of energy into the ICM.
This event triggers the formation of a wealth  of dense blobs. It is
important to realize that these blobs are non-linear perturbations of the ICM,
and may be distributed with a wide spectrum of densities. These blobs
are denser than the surrounding medium, and fall to the black hole.
If these blobs have an initial angular momentum,
they do not freely fall, but may form an accretion disc. As we shall
demonstrate (\S~\ref{s:am}),  for the expected angular momentum 
distribution this is not a
relevant complication. The dense and cool blobs are the fastest to fall,
and therefore  are removed first from the ICM. The accretion
history may be difficult to predict, since it depends on several factors.
For most of the time the small blobs  may accrete  approximately in a
steady-state, but some processes may intervene to modify this.
For instance if most of the blobs are formed at the same distance to the
AGN with a  sharply peaked density spectrum,    most of the blobs might
accrete simultaneously.  Moreover, an accretion disc may form and
undergo some kind of instability. In either case, a sudden  ``catastrophic'' 
accretion episode
on the black hole is expected, resulting in a new AGN outburst, which restarts
the cycle with a fresh injection of  blobs.
We examine here  below some  details of the suggested process.

In this duty cycle 
some of the gas cools to low temperatures ($\lesssim 10^4 \K$) 
before the next major heating, while the rest is heated back to
a relatively high temperature.
We differ from many previous models in that in the moderate CF model
a substantial fraction of the ICM gas cools to low temperatures.
The presence of a detectable amount of gas cooling  below X-ray
emitting temperatures  is a prediction of this model.
Indeed, in the CF cluster Abell~2597 both extreme-UV and X-ray
observations indicate a mass cooling rate of $\sim 100 \msyr$,
which is $\sim 0.2$ of the value quoted in the past based on {\sl ROSAT} 
X-ray observations \citep[see the discussion in][]{Mor05}.
In the CF cluster Abell~2029, \citet{Cla04} find a substantial
amount of gas at a temperature of $\approx 10^6 \K$; a CF model gives
a mass cooling rate of $\sim 50 \msyr$.

The AGN outburst interacts in a very complicated fashion with the ICM
\citep[e.g.][]{Beg04}, e.g. it heats and inflates radio bubbles,
which rise buoyantly in the ICM.
The ICM itself is displaced and thickened by the rising bubbles,
as shown by the enhanced X-ray brightness \citep[e.g.][]{Bla01}.
A non-homogeneous thickening  may result in the formation of a 
multi-phase  gas inside or near the radio lobes, which harbor 
relatively strong  magnetic fields, up to few tens $\mu G$.

The magnetic field inside the radio lobes is related to a fundamental 
issue for the cold feedback model, namely the efficiency of thermal
conduction. A highly efficient conduction would evaporate the 
cold gas blobs  before they can accrete on the AGN, and is therefore 
incompatible with the cold feedback model. 

It is well-known that magnetic fields are able to suppress
thermal conduction, and that the degree of suppression strongly 
depends on their  topology
\citep[see e.g. the discussion in][]{Nar01}. 

For our purposes, we assume that there is essentially {\em no}  
heat conduction between the over-dense blobs and their surroundings,
i.e., the over-dense blobs are magnetically disconnected from their 
environment,
as in the model of \citet{Sok04a}.  Effectively, for our  blobs of radius 
$a \sim 10-100 \pc$,
we find from Figure~3 of \citet{Nip04}
that the effective heat conduction should be
$\la 0.001$ times the \citet{Spi56} value in order for
the blobs not to be evaporated.

This suppression factor  is somewhat high, but   consistent with some 
recent observations. \citet{Mol02}, e.g., 
finds evidence of a gas component cooler than the ambient gas
inside the radio lobes of M87. The very existence of these cold pockets
led \citet{Mol02} to estimate a conduction suppression  
factor~$\la 0.01$ with respect to the nominal \citet{Spi56} value.

The conclusion is that  the radio lobes, permeated by magnetic fields, may be
``safe corridors''  where the blobs may accrete on the AGN without
being evaporated by  conduction.


The coexistence of gas phases at different temperatures is not exclusive of
M87. In their analysis of the group NGC~5044, \citet{Buo03} find  evidence
for a moderate multi-phaseness.
Their data are well fit by a two temperature model, a relatively cool 
component with $T_{\rm cool}\approx 0.7$~keV and a hot
component at $T_{\rm hot}\approx 1.4$~keV. These temperatures seem to
coexist in the inner $\approx 30$~kpc, with the cooler component dominating
in the inner $\approx 10$~kpc. The  cool component has a sizeable filling
factor:  $f_B \approx 0.5$ for $r\lesssim 25$~kpc and $f_B \approx 0.1$
at  $r\gtrsim 30$~kpc.

We therefore consider the inner region, $r \lesssim 5-30 \kpc$ of
cooling-flow clusters to
posses an ICM with non-linear perturbations, i.e.
dense blobs spread within it.
The fate of a blob depends on the relative magnitudes  of its
cooling time $t_{\rm cool}$,  the time interval to the next AGN heating event,
and on  the time $t_{\rm fall}$ the blob takes to accrete on the central black
hole. The time scale  $t_{\rm fall}$ depends on several factors. The first is
the blob's over-density with respect to the ambient gas. The second is the
blob's angular momentum: a high angular momentum prevents the blob from
approaching the black hole altogether.
We shall discuss the issue of angular
momentum more fully in \S~\ref{s:am}: for the time being we
assume that the content of angular momentum of a blob is small.
Under this hypothesis, the relation
between  the time scales  $t_{\rm cool}$ and $t_{\rm fall}$
mainly depends on the blob's density  $\rho$
relative to that of the ambient medium $\rho_a$.
A sinking blob has $t_{\rm cool} \lesssim  t_{\rm fall}$, with the extreme
case of $t_{\rm cool} \ll  t_{\rm fall}$, where the blob
significantly cools down on a very short time scale, and almost  free-falls
to the center.
If $t_{\rm fall} < t_{\rm cool}$ the blob would not cool much, and
will reach a position where it is  as dense as the ambient gas; if
$t_{\rm cool} \approx  t_{\rm fall}$ the blob sinks fairly slowly,
owing to its small over-density.

In this way, the ICM efficiently disposes of the cooler phases, i.e.,
the highly non-linear over-dense perturbations.

If a blob cools isobarically in pressure equilibrium with its
surrounding, its cooling time scales as $t_{\rm cool}\propto \rho^{-2}$.
Later on, when the blob temperature drops below $T\approx 0.1 \keV$,
the sound waves become too slow to keep up with the outer pressure,
and cooling occurs isochorically: $t_{\rm cool}\propto \rho^{-1}$
\citep[e.g.][]{Bur00}.
In either case  the denser  ---~and the cooler~--- is  a perturbation,
the faster  it cools, and the more efficiently it is removed from the ICM.

The discussion above leads us to the proposed scenario.
Non-linear over-dense blobs of gas, $\delta \rho /\rho_a \gtrsim 2$,
i.e.  $\rho/\rho_a \gtrsim 3$, cool on short time scales such that
they are removed from the ICM before the next major AGN heating event.
Some of these blobs cool and sink toward the central black hole.
Other non-linear perturbations  may form stars, as is inferred in some
CF clusters, e.g., in the CF cluster A1068 the cooling rate within
$r \approx 30 \kpc$ is about equal to the star formation rate there
\citep[][]{Wis04, McN04}.
The dense blobs that sink to the center feed the AGN.
The feedback is with the {\it  entire} cool inner region, and not
only with the gas close to the black hole.
Any over-cooling taking place  in the inner region, where the
temperature profile is flat, will lead to many small and dense blobs, which
feed the AGN.

We assume that the cold blobs are magnetically disconnected from 
the surrounding, so that most of them  can survive long enough 
to be delivered to the central AGN.  Indeed, while some blobs
might certainly be evaporated,  yet the inability of
thermal conduction to re-heat {\em all} the cold gas is testified 
by the presence of sizeable amounts of molecular gas in the 
central few kpc of some clusters (NGC~1275/Perseus: \citealp{Ino96}).
In a recent paper  \citet{Wil05} resolved a ring
of molecular gas with radius of $50 \pc$ from the center of NGC~1275 at the
center of the Perseus  cluster. Adding the presence of molecular gas
at distances of up to $\sim 10 \kpc$, their finding shows that cold
gas originating at large distance in the cluster, but still within the low
temperature region,  can feed the central black hole.

This ends our simple demonstration that highly over-dense blobs can
be accreted to the central black hole before the next major AGN heating
event.
More detailed  calculations are presented in the next Section;
readers not interested in them may  skip directly to
the last Section.

\section{Nonlinear Evolution of the Blobs}
\label{s:nnlin}

In this Section we consider the evolution of a single blob. Let
$V$ and $S$ be its volume and cross-section, respectively, and
$\rho$ its mass density. The blob is subjected to the overall gravitational
acceleration $\g$, to the hydrostatic buoyancy force and to the drag force.
Its
equation of motion may be written \citep[e.g.][]{Loe89, Kai03}
\be
\label{e:motion1}
\rho\:  V\:  \dtot{\v}{t} = \g\;(\rho - \rho_a) \: V
- \frac{C}{2}\: S\: \rho_a\: v\: \v,
\ee
where $\rho_a$ is the mass  density of the ambient gas and $C$ is the
dimensionless drag coefficient. For $C$ we shall adopt the value $C\approx 0.75$
used by \citet{Kai03} and derived from the numerical simulations by
\citet{Chu01}.
It will be convenient to express Equation~(\ref{e:motion1}) in terms of the
blob over-density $\delta$ with respect to the surrounding medium
\be
\label{e:overd}
\delta \equiv \frac{\rho - \rho_a}{\rho_a}.
\ee
Eliminating $\rho$ from Equation~(\ref{e:motion1}), we may rewrite it as
\be
\label{e:motion}
\dtot{\v}{t} = \g\; \frac{\delta}{1 + \delta}
- \frac{3 C}{8}\: \frac{v}{1 + \delta}\:  \frac{\v}{a},
\ee
where we have calculated $V$ and $S$ assuming that the blob is a
sphere of radius $a$.
If $a$, $\delta$ and $g$ are constants, the last equation is easily integrated:
\be
\label{e:velo}
v = v_t \;  \frac{e^{t/\tau} - 1}{e^{t/\tau} + 1},
\ee
where
\be
\label{e:tau}
\tau = \left(\frac{2\: a}{3\: C\: g}\right)^{1/2} \;
\frac{1 + \delta}{\delta^{1/2}}
\approx
5 \times 10^6  \left(\frac{a}{100\pc}\right)^{1/2}  \left(\frac{g}{10^{-8}\cm\:{\rm s}^{-2}}\right)^{-1/2}  \;   \frac{1 + \delta}{\delta^{1/2}}\; \yr
\ee
is the characteristic time in which the blob attains its terminal velocity
\be
\label{e:vterm}
v_t = \left(\frac{8}{3\,C}\:  g \: \: a\; \delta \right)^{1/2} \approx
33 \left(\frac{a}{100\pc}\right)^{1/2}  \left(\frac{g}{10^{-8}\cm\:{\rm s}^{-2}}\right)^{1/2}  \; \delta^{1/2}\;  \km\; {\rm s}^{-1}.
\ee
Note that if the gravitational acceleration $g$ is relatively small,
then a blob will take more time $\tau\propto g^{-1/2}$ to accelerate to a
smaller  terminal velocity $v_t\propto g^{1/2}$. This explains why in
Figure~\ref{f:times} the terminal-velocity fall time drops below the
free-fall time at small radii, where $g$ is small;
the blobs starting there take a comparatively long time to accelerate.

In order to calculate the evolution of a blob, we need three
further equations.
The first  equation simply  relates the position and the velocity of the blob:
\be
\label{e:rv}
\dtot{\bm r}{t} = \v.
\ee
The second equation is the  blob mass conservation
\be
\label{e:mass}
\rho\: a^3 ={\; \rm constant}.
\ee
The last equation provides  the blob  energy balance.
If $E$ is the blob total internal energy and $P$ is its pressure,
the combined first and second law of thermodynamics
\be
\label{e:blob1}
P \; d S = - n_e\, n_H\,\Lambda\, V\: dt
\ee
yield the variation  of the blob entropy
\be
\label{e:entropy}
S = \frac{3}{2}\; \ln \left(\frac{P}{n^{5/3}}\right)
\ee
under the radiative loss  of the amount
of energy $d Q =  n_e\, n_H\,\Lambda\, V\: dt$ in the time interval $dt$.
This loss  depends on the cooling function $\Lambda(T)$ and
on the electron and hydrogen densities  $n_e$ and $n_H$ of the gas inside
the blob.
We now suppose that the blob is instant by instant in pressure equilibrium
with the ambient gas, whose pressure is $P_a$:
\be
P \equiv P_a.
\ee
This assumption holds  until
the blob sound crossing time is short in comparison to the blob cooling time,
and  therefore the  acoustic waves can efficiently pressurize the blob
\citep[see e.g.][]{Bur00}.
We shall have to check {\it a posteriori} the validity of this hypothesis.
We further assume that the ambient gas is in hydrostatic equilibrium.
With these hypotheses, the pressure variation $d P$ experienced by the blob is
only due to its drift across the cluster gravitational potential $\Phi$:
\be
d P \equiv  d P_a = -\rho_a \; d \Phi =  -\rho_a \; dt \: \v \cdot \nabla  \Phi
\ee
With the help of this equation we differentiate Equation~(\ref{e:entropy})
and plug the result into Equation~(\ref{e:blob1}): after some
straightforward algebra we obtain
\be
\label{e:ndot}
\dtot{\, \ln n}{\,t} =  \frac{n_a}{n} \; \frac{\g\cdot\v}{c_s^2}
+  \frac{2}{5}\; \frac{n_e\: n_H}{n} \: \frac{\Lambda(T)}{k T},
\ee
where $c_s^2 = 5\, k T / 3 \, \mu m_p$
is the (squared) adiabatic sound speed inside the blob, $k$ is the Boltzmann
constant,
$\g = -\nabla\Phi$ is the  gravitational acceleration
and $n_a=\rho_a/\mu m_p$ is the ambient  total number particle density.

Using the  definition~\ref{e:overd}, we may rewrite Equation~(\ref{e:ndot})
in terms of the over-density $\delta$:
\be
\label{e:radiat}
\dtot{\delta}{t} = \frac{\g\cdot\v}{c_s^2}
+ \frac{2}{5}\;  \mu_e\: \mu_H\: n_a \; (1 + \delta)^2 \; \frac{\Lambda(T)}{k T} - (1 + \delta)\;  \v\cdot\nabla\ln n_a,
\ee
where the coefficients  $\mu_e$ and  $\mu_H$ are defined by
$\mu_e \equiv n_e/n$ and $\mu_H \equiv n_H/n$. The first term
on the  right-hand side of Equation~(\ref{e:radiat})
is the  adiabatic compression owing to gravity;
the second term describes the blob thickening at constant pressure as its
temperature cools on account of the radiative losses; the last term
is proportional to the slope on the background ambient gas, and is
the reduction of the density contrast owing to the denser layers the
blob  sinks through. This last term is important in what it may
hamper the growth of large over-densities in a strongly stratified ambient gas.

Equations~(\ref{e:overd},~\ref{e:motion},~\ref{e:rv},~\ref{e:mass})
and~(\ref{e:radiat})
provide a closed system of equations for the blob evolution,  once we have
specified the  distribution of the ambient density $\rho_a$,
the cluster gravitational potential $\Phi$ and a suitable set of initial
conditions.

It is important to remark that in our derivation we have assumed that the
ambient gas behaves as a static background without cooling. This approximation
is justified as far as the blobs cool faster than the ambient gas.
This assumption certainly holds  if the blob remains  over-dense, but may
break down if  $|\delta| \ll 1$ \citep[][]{Loe89}.
Therefore, our results concerning blobs
only slightly over-dense with respect to their surroundings require some care.

\bigskip

We apply the above  equations  to calculate the evolution of a blob in the
cool core cluster  M87. We choose this cluster because we know to a
good approximation its
gravitational mass, its temperature and  density structure;
besides, the central AGN here shows a complex interaction with the
surrounding medium \citep[e.g.][]{Bel01, Mol02}.
Where necessary, however, we shall extend our considerations
beyond the particular features of M87.
For the ambient density $\rho_a$, temperature $T_a$ and
gravitational acceleration  $\g$ of M87 we assume the functional forms
provided by  \citet{Ghi04}.
We have employed the cooling function $\Lambda(T)$ given  by
\citet{Sut93} for a solar-abundance thermal plasma.
As far as the initial conditions, we assume that the  blob has an
initial over-density $\delta_0 > 0$, and is
released from rest at the  distance  $r_0$ from the cluster center.
We have solved numerically the system of
Equations(~\ref{e:overd},~\ref{e:motion},~\ref{e:mass},~\ref{e:radiat},~\ref{e:rv}) and these initial conditions with a step-adaptive fifth-order
Runge-Kutta algorithm \citep[][]{Pre92}.

Figure~\ref{f:ddd} plots the  over-density evolution of some blobs, which
only  differ in their  initial over-densities $\delta_0$;
all of them have initial size  $a_0=100$~pc, and have been
released from rest at  $r_0=20$~kpc from the cluster center.
It is apparent that the qualitative behavior of the solution critically
depends on $\delta_0$: if  $\delta_0$ is below a
critical  threshold $\delta_C$ the over-density  decreases, else if
$\delta_0 > \delta_C$ it diverges.
A blob with $\delta_0 < \delta_C$  soon evolves to   $|\delta|\ll 1$,
becoming hardly distinguishable from the  ambient background gas, whose
cooling and bulk velocity have been neglected in our approximation.
Therefore, our result concerning these blobs may be looked at suspiciously,
and for consistency we conservatively disregard their
contribution to the feeding of the central black hole.
These blobs are most likely to get pinned by the drag force
to the average bulk motion  of the ICM, and their successive evolution
would require a more detailed analysis
\citep[see e.g.][]{Nul86, Loe89}.
To our concern, since the background medium has been assumed static,
we may assume  that such  blobs have been thermally  stabilized.

The existence of a critical over-density  $\delta_C$
 is not difficult to understand \citep[][]{Bal89, Loe89,
Tri91}.
As a blob sinks, it crosses thicker and thicker  layers of ambient gas, and
its density contrast $\delta$ with respect to its surroundings would reduce.
If this effect is not contrasted, the blob would stop its fall at an
equilibrium  distance
from the center where its density is the same as the ambient gas, i.e.
$\delta=0$.
The only way to overcome this effect is to cool fast. Since this
process is isobaric, the blob density is enhanced, and  the blob
might reach the cluster center {\it before} its over-density has
significantly reduced. Since the blob cools on a time-scale
$t_{\rm cool}$,  a blob  condenses and reaches the center if
$t_{\rm cool} \lesssim t_{\rm fall}$. The equality between these time-scales
defines the  critical value $\delta_C$ for the blob over-density.
This explains the  qualitative behavior shown in Figure~\ref{f:ddd}.

Figure~\ref{f:rv} plots the evolution of the 
position and velocity of two blobs, one slightly
over-critical ($\delta \gtrsim \delta_C$)  and one slightly under-critical
($\delta \lesssim \delta_C$).
In both cases, the  blob is initially in  free-fall, but after a
transition time of the order of  few $\tau$ --given by Equation~(\ref{e:tau})--
the drag limits the velocity to the  terminal value $v_t$ given by
Equation~(\ref{e:vterm}).
An under-critical blob attains  with non-zero velocity its
equilibrium position  (where $\delta=0$), which is
overshot. The blob starts a series to  oscillations  about this position,
which are quickly  damped by  the drag force
\citep[e.g.][]{Bal89, Loe89, Tri91}.
Eventually, the blob cools and becomes  pinned to the bulk  of
the ambient gas.
An over-critical blob  is always denser than the surrounding gas,
the drag plays a minor role, and the blob nearly free falls all the way
down to the center.

The evolution of the  blob temperature is presented in Figure~\ref{f:temp}.
The under-critical blob has  an initial radiative cooling time of
$t_{\rm cool}\approx 1.8\times 10^8$~yr, and before this time the blob
temperature does not change much. For such a blob  the
gravitational heating is  more important than  the radiative losses:
the blob is heated up to the local ambient temperature, and
becomes  thermally stable.
In a denser blob, on the contrary, the gravitational heating is insufficient
to overcome the radiative cooling.
After a time  $t_{\rm cool}\approx 1.2\times 10^8$~yr,
therefore, the blob cools very fast and almost free falls to the cluster
center.

The initial release radius is important for the nonlinear development
of  a blob.  Figure~\ref{f:radii} plots the over-density evolution for some
blobs with the same intrinsic properties, but  released at different
radii. It is apparent that the farthest blobs never develop large
over-densities. This fact is related to the density profile of the ambient gas
(shown in Figure~\ref{f:gas}). For a fixed  over-density, a far blob
has a relatively small density --in absolute terms-- and as it sinks it
soon gets embedded in the denser ambient gas: the over-density reduces
and the blob is  thermally stabilized.
The situation is different if the blob  has been released at a small distance
to the center, where the ambient density profile is flat
(Figure~\ref{f:gas}). As in our discussion following
Equation~(\ref{e:radiat}),
in this case as the blob's over-density growth
is not hindered  by the ambient gas gradient,  and even a moderate initial
over-density may evolve towards larger values.
It is worth to stress the importance of this density stratification  effect,
as  it relates the size of the
central density plateau to the amount of allowed cold feedback in our model.
Indeed, in the density plateau  essentially all the over-densities may
evolve to the non-linear regime, cool down and accrete on the AGN to
provide the heating feedback.
On the other hand, only the largest over-densities born outside the
central plateau may evolve this way. In the case of M87 this
plateau extends out only  to $r\approx 5$~kpc (Figure~\ref{f:gas}):
more typical cooling flow clusters, like A2052, have  $r\approx 30$~kpc
\citep[][]{Bla01}; this  allows the feedback between the AGN and  a
larger  fraction of the  cluster's gas.

Following \citet{Tri91} we  estimate the
critical over-density $\delta_C$ as a function of the other parameters.
Equation~(\ref{e:radiat}) evaluated at $t=0$ with our initial
conditions yields
$\dot{\delta}>0$. If the blob is stable, this initial
trend must be reversed, and   there must be
an instant $t=\tilde{t}$ for which $\dot{\delta}(t=\tilde{t})=0$. If the blob
is unstable, on the other hand,  $\dot{\delta}>0$ all the way.
Equation~(\ref{e:radiat}) evaluated at $t=\tilde{t}$ may be written
\be
\label{e:dcrit}
(1 + \delta)^2  = \left| \frac{g r}{c_a^2} +  \dtot{\,\ln n_a}{\,\ln r}\right| \frac{t_a \: v}{r} \;  \frac{\Lambda(T_a)}{\Lambda(T)},
\ee
where $c_a^2 = 5\, k T_a/\, 3\mu m_p$ and
$t_a = 5\, k T_a/ 2\,\mu_e\mu_H n_a  \Lambda(T_a)$
are respectively the sound speed and the isobaric cooling time of the
ambient gas; in pressure equilibrium the blob temperature $T$ and the
ambient gas  temperature $T_a$ are related by $T =T_a/(1 + \delta)$.
We must evaluate the right-hand side of Equation~(\ref{e:dcrit}) at
$t=\tilde{t}$. If $\tau \ll \tilde{t}$ (where $\tau$ is given by
Equation~\ref{e:tau}), we may substitute $v$ with its terminal
value given by Equation~(\ref{e:vterm}). If we suppose that the  other 
quantities
on the right-hand side of Equation~(\ref{e:dcrit}) are not too different 
from their initial values, the last equation may be rewritten
\be
\frac{(1 + \delta)^2}{\delta^{1/2}} = t_a\; \left(\frac{8}{3\,C}\; g \;r \right)^{1/2} \left|\frac{g\,r}{c_a^2} + \dtot{\,\ln n_a}{\,\ln r}\right| \; \left(\frac{a}{r}\right)^{1/2} \: \frac{\Lambda(T_a)}{\Lambda(T)}.
\ee
The value of $\delta$ provided by this equation is the critical threshold
$\delta_C$ between a thermally stable and a thermally unstable blob.
In the non-linear regime $\delta\gg 1$, by approximating
the cooling function  by a power law $\Lambda \propto T^\alpha$,
we obtain $\delta_C\propto a^{1/(3-2\alpha)}$,
which  coincides with the expression  given by \citet{Tri91} if
$\alpha=1/2$, i.e. if thermal Bremsstrahlung is the chief coolant.
Figure~\ref{f:deltacrit} plots $\delta_C$ as a function of $a$ for
different  release radii $r$ \citep[see also the  Figure~1 of][]{Tri91}.
In the upper branch of the $\delta_C-a$~plot we find
$\delta_C\propto a^{1/3}$ owing to the weak dependence of the cooling
function on $T$ in the temperature range typical  of M87.
It is interesting to note that the over-density required for a blob
to be  unstable   is moderate: $\delta_C\approx 0.8-3$ if the blobs have sizes
$a\approx 10-100$~pc. From Figure~\ref{f:deltacrit} we also note that
the blobs have a  minimum stable size. As it is seen from
Equation~(\ref{e:radiat}), the heating term is proportional to the fall
velocity,
which is $v_t\propto a^{1/2}$, where $a$ is the blob radius. On the other hand,
the radiative cooling term does not depend on the blob size. Therefore, we
expect that if a blob is too small (i.e. below a critical threshold size)
the gravitational term is insufficient to heat the blob:  the  radiative
cooling prevails altogether, and the blob's over-density increases
monotonically with time.

Before ending this Section, we must notice that for the very over-dense blobs
the assumption of
isobaric cooling may break down, since the sound waves inside the blob
become too slow to pressurize it against the ambient gas. The evolution is
now isochoric, and the blob density  grows less than the amount
predicted by our isobaric  model \citep[e.g.][]{Bur00}.
If this transition occurs before the blob has reached the inner density
plateau, its density could be not high enough to avoid the thermal
stabilization. In a more typical situation the
transition to the isochoric regime occurs when the blob has already reached a
sizeable over-density, so this effect is generally not very important,   and
the qualitative  conclusions drawn under the isobaric
assumption are not altered; the quantitative results may be different
by small amounts for the small  blobs.





\subsection{The Issue of Angular Momentum}
\label{s:am}

In our qualitative sketch we have so far omitted any reference to the
angular momentum of the infalling material.
A too large angular momentum might prevent the flow from
approaching the central black hole: the flux would merely stagnate,
cool down and condense in filaments or stars.
The AGN fueling is cut off altogether, which  makes the feedback impossible
\citep[see][for a thorough analysis of the thermal instability of a
high angular momentum flow]{Cow80}.
The existence of a circumnuclear disc around M87 shows that the
the flux possesses an amount of angular momentum, so the question is
whether  this angular momentum is high enough to jeopardize the
feedback. We argue it is not.

First of all, as we will discuss below, the blobs are expected to 
form and accrete only in a region of the same extension as the inner
gas density plateau ($5-30\kpc$). The distribution of angular 
momentum at larger radii is immaterial to the present discussion, and
in the rest of this Section we only refer to the ICM within this central 
region.

The non-linear perturbation  spectrum may stem  directly from  ICM
disturbances  driven by an early AGN activity,
but also  from galaxies mass-stripping \citep[][]{Sok91}.
Since the  galaxies do not have an ordered bulk motion, the blobs
stripped from them are  also unlikely to  organize in an
ordered  flow with high net angular momentum.
Besides, if a blob from a galaxy is injected with a high angular
momentum, it is likely to loose most of it on account of  its  friction with 
the surrounding ICM. Therefore, even if a circular flow like a disc may form,
it cannot be very large, as the example of M87 shows.

A flow with specific angular momentum $l$ circularizes  at the  radius
$R_{\rm circ} \approx  l^2/G\,M_{\rm BH}$, being   $M_{\rm BH}$ the central 
black hole mass.
We estimate $R_{\rm circ}$ as follows.  Let us consider a cold blob of
radius $a$ initially at the
distance $R$ from the central black hole.
The balance between gravitational attraction and  the friction force
quickly brings the blob to the terminal velocity $v_t$ given by
Equation~(\ref{e:vterm}). Then, its orbital angular momentum is about
$l \approx v_t\: R$ and  Equation~(\ref{e:vterm})
gives $R_{\rm circ} \approx a\:\delta$.
The  circularization radius is expected of the same size as the non-linear
density  blobs,  since $\delta$ is expected to be of the order of a few.
Rough as it is, this estimate is
in fair agreement with the actual size of the circumnuclear
disc of M87 $R_d \approx 10^2\pc$ \citep[][see also below]{Har94, For94}.
We note that, statistically, many of the cold blobs will
start with very low angular momentum $l \ll v_t\:  R$. These
will be accreted directly to the black hole vicinity.

\bigskip

As an example we consider the CF cluster M87.
First we note that M87 has a large Keplerian disc at its center.
\citet{Har94} estimate the black hole mass to
be $M_{\rm BH}\approx 2.4 \times 10^9 \msun$ from their observation
of a disc with a radius of $\approx 20 \pc$.
Their optical {\sl HST} image shows the disc to be
$\approx 3.5$ times larger, i.e., $R_d \approx 70 \pc$
\citep[][]{For94}.

The cooler gas at the center of the Virgo cluster is at
a temperature of $\approx 1 \keV$, which implies for the
above black hole mass a Bondi accretion radius of
$\approx 80\pc$ \citep[][]{Chu02}
\footnote{
\citet{Mac97} estimate a somewhat larger mass for the central black
hole in M87, namely
$M_{\rm BH} \approx (3.2 \pm 0.9) \times 10^9 \msun$.
In  this case, the
Bondi accretion radius is slightly larger, about $100 \pc$.}.
A Bondi accretion radius  as large   as  the
disc around the black hole of M87  further suggests that
the simple Bondi accretion flow
\citep[][]{Chu02, Nul04} does not hold; the accreted material
has a larger angular momentum, and may come from much larger radii.

To further elaborate on the proposed model, we plot in Figure~\ref{f:times}
the free fall time in M87, as well as some relevant cooling times.
In that Figure the crosses  mark the cooling time
of the ambient gas at several radii; the asterisks, the filled
and the empty  triangles
are the cooling times of  blobs  with local over-densities with respect
to the ambient of  $\delta\rho/\rho_a=1$,  $\delta\rho/\rho_a=3$ and
$\delta\rho/\rho_a=10$, respectively.
Pressure  equilibrium  is assumed between the blobs and the ambient medium:
since the  blobs temperature is always above $T\approx 0.1$~keV,
the sound waves are fast enough to ensure this equilibrium.
The solid line represents the free fall time in the gravitational potential
of M87; the dashed line is the fall time of a blob of radius $a=300$~pc
and over-density  $\delta\rho/\rho_a=1$.
In this case, the fall velocity is given by the balance between the
gravitational acceleration and the drag force.
The actual fall time lies in  between these
time-scales (see the next Section).
The blobs for which $t_{\rm cool} \approx  t_{\rm fall}$ have moderate
over-densities $\delta\rho/\rho_a=1-2$, and therefore temperatures of
$1/2-1/3$ of the ambient gas.
This value is not far from the values found by \citet{Mol02}
for the cool component in M87, and in good agreement with
the cool component of NGC~5044 \citep[][]{Buo03}.


An important  final  remark is in order. As \citet{Nul86} pointed out,
{\it in the absence of a cohesive force}
a blob would be torn apart  by  the ram pressure
in a characteristic time
\be
t_{d} = \frac{a}{v_t} \left(\frac{\rho}{\rho_a}\right)^{1/2}
\approx 10^7\;
\left(\frac{a}{100\pc}\right)\:
\left(\frac{v_t}{10\kms}\right)^{-1}
\left(\frac{\rho}{\rho_a}\right)^{1/2}\; \yr,
\ee
where $a$ is the blob's radius and $v_t$ its terminal velocity,
given by  Equation~(\ref{e:vterm}). This may be considerably shorter
than the time taken by the blob to fall to the center, and we must therefore
assume that some kind of cohesive force (like a magnetic tension) is at work
to prevent the blob disruption. In any case, the proposed scenario works also
for the smaller blobs forming from the fragmentation of a larger blob.

\section{Summary}
\label{s:summary}

This paper deals with heating the intra-cluster medium (ICM) in
galactic and cluster cooling flows (CF) by an active galactic
nucleus (AGN) sitting at the cooling flow center.
As was shown by many papers, the heating is most likely to take place
via a feedback mechanism, where the ICM cooling enhances the
AGN activity, which in turn heats the ICM and  quenches the cooling flow.

\begin{figure}
\centering
\includegraphics[width =85mm, angle = -90]{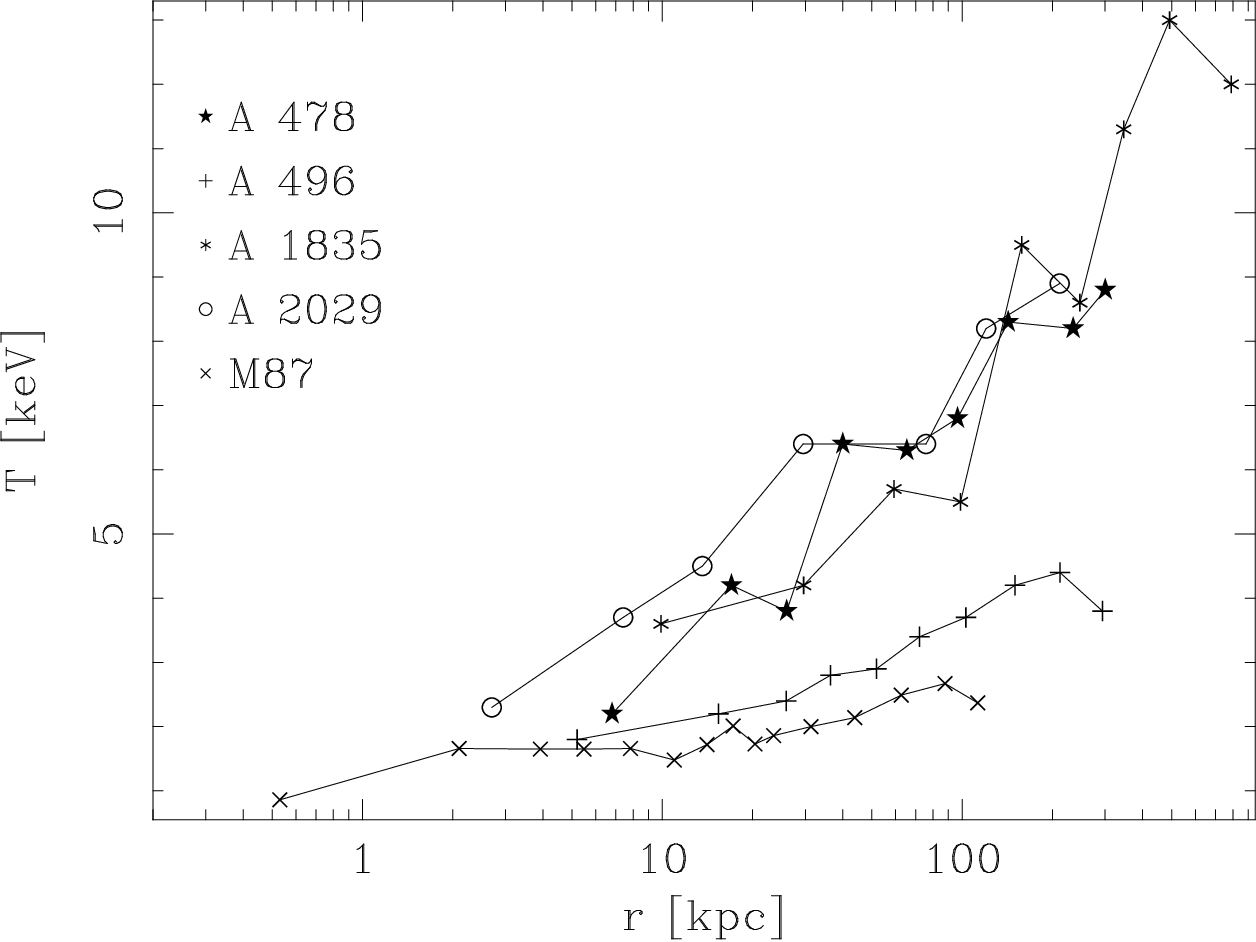}
\includegraphics[width =85mm, angle = -90]{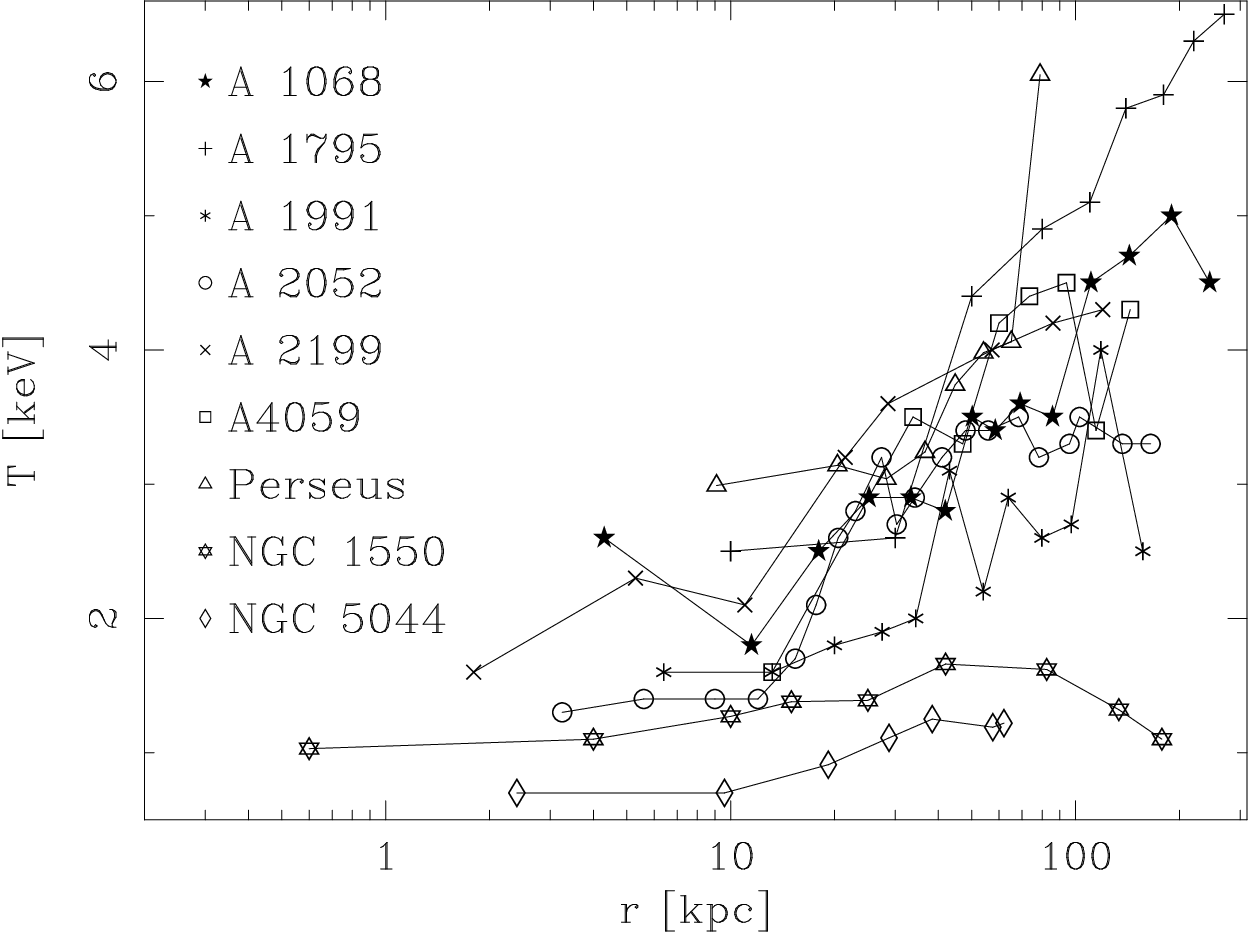}
\caption{\label{f:tprof}
The temperature profiles for a sample of Clusters observed with
{\sl Chandra} (C) or {\sl XMM-Newton} (X):
A478 \citep[][C]{Sun03a},
A496 \citep[][X]{Tam01},
A1068 \citep[][C]{Wis04},
A1795 \citep[][C]{Ett02},
A1835 \citep[][C]{Sch01},
A1991 \citep[][C]{Sha04},
A2029 \citep[][C]{Lew03},
A2052 \citep[][C]{Bla01},
A2199 \citep[][C]{Joh02},
A4059 \citep[][C]{Cho04},
M87 \citep[][C+X]{Ghi04},
Perseus  \citep[][C]{Sch02},
and finally the groups
NGC~1550 \citep[][C]{Sun03b} and
NGC~5044 \citep[][C]{Buo03}.
For the sake of readability, the clusters with an approximatively
power-law temperature profiles and those with a central
temperature floor have been
plotted in two different panels (upper and lower, respectively).
Where necessary, the radii from the original papers have been
corrected for a cosmology with $H_0~=~70$~km/s/Mpc.
}
\end{figure}

\begin{figure}
\centering
\includegraphics[width =85mm, angle = -90]{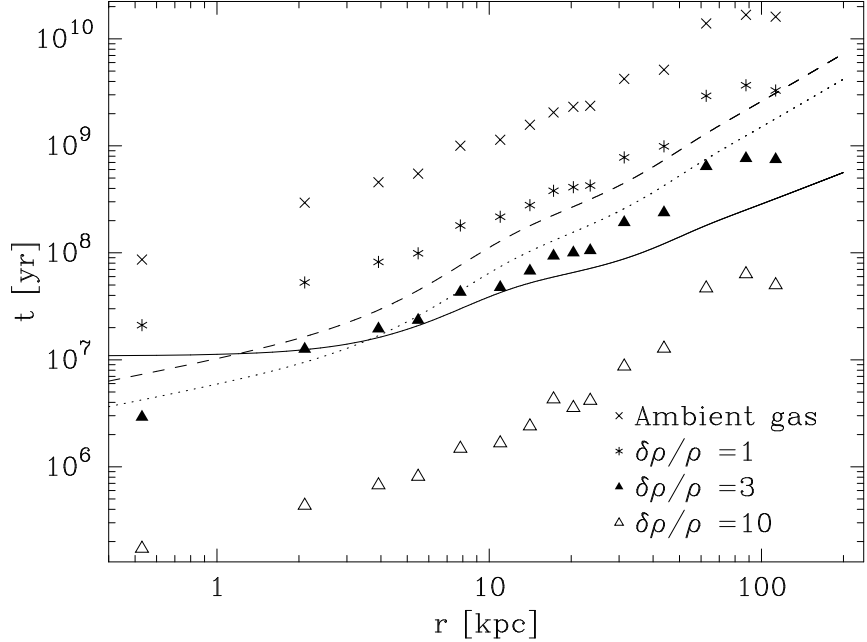}
\caption{\label{f:times}
A comparison between the free fall time, the terminal-velocity fall time
and the isobaric cooling times for M87. The free fall time scale
$t_{\rm ff} \approx (2\, r / g )^{1/2}$ is shown by the solid line;
the terminal-velocity  fall times $t_{t} \approx r / v_{t}$
refer to  a blob of radius $a=300$~pc and an over-density
$\delta \rho/\rho_a =1$ (dashed line) and $\delta \rho/\rho_a =3$
(dotted line). The gravitational acceleration  has been
calculated from the deprojected density and temperature profiles, with the
additional hypothesis of hydrostatic equilibrium. For the data and the
procedure, see \citet{Ghi04}, and references therein.
The cooling  times have been
calculated with the deprojected data of \citet{Ghi04}, assuming an
average metal  abundance of $Z/Z_\odot=1.0$
\citep[expressed in][solar units]{And89}, with the cooling function
provided by \citet{Sut93}.
The crosses  refer to the ambient gas, asterisks to  gas blobs with
local  over-densities  $\delta\rho/\rho_a =1$,  the filled and empty
triangles respectively to  gas blobs with
local  over-densities  $\delta\rho/\rho_a =3$ and $\delta\rho/\rho_a =10$
with respect to the ambient gas. Pressure equilibrium between  the blobs
and the ambient is assumed.
The  plot does not extend beyond  $0.4\kpc$ because the density
and temperature profiles we have taken from \citet{Ghi04} and that we 
used to calculate the characteristic time scales do not push to 
smaller radii.}
\end{figure}

\begin{figure}
\centering
\includegraphics[width =85mm, angle = -90]{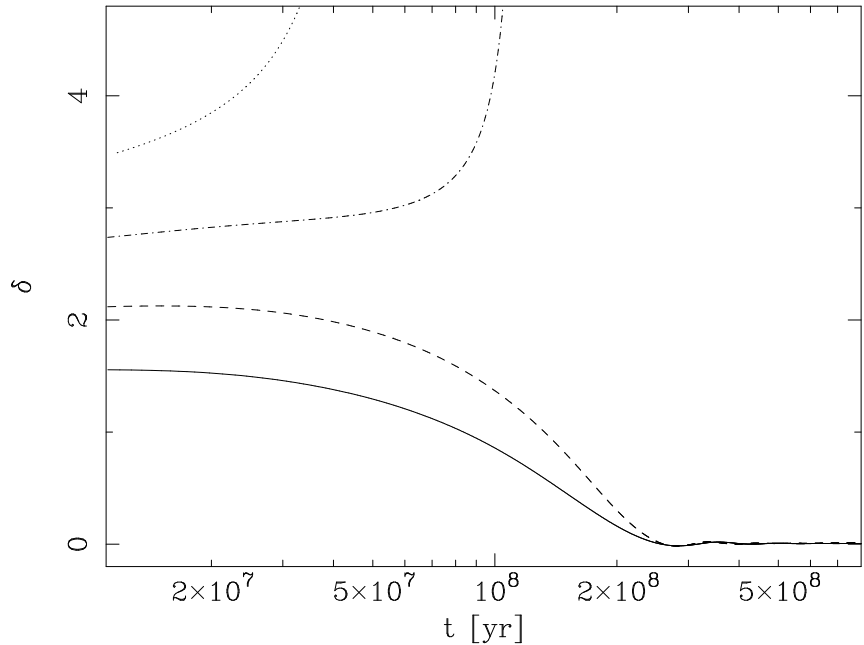}
\caption{\label{f:ddd}
The evolution of blobs with $a_0=100$~pc, $r_0=20$~kpc, but different \
initial   over-densities $\delta_0 = (\rho -\rho_a)/\rho_a$:
$\delta_0=1.5$ (solid line), $\delta_0=2.0$ (dashed line),
$\delta_0=2.5$ (dot-dashed line), $\delta_0=3.0$ (dotted line). Below a
critical threshold $\delta_C$ of the initial over-density $\delta_0$
(in this case $2<\delta_C<2.5$), the blobs
are stabilized, above this limit they condense in a short (cooling) time.
Here and in the following Figures the ICM properties reproduce those of M87,
whose data have been derived from \citet{Ghi04}.
}
\end{figure}

\begin{figure}
\centering
\includegraphics[width =85mm, angle = -90]{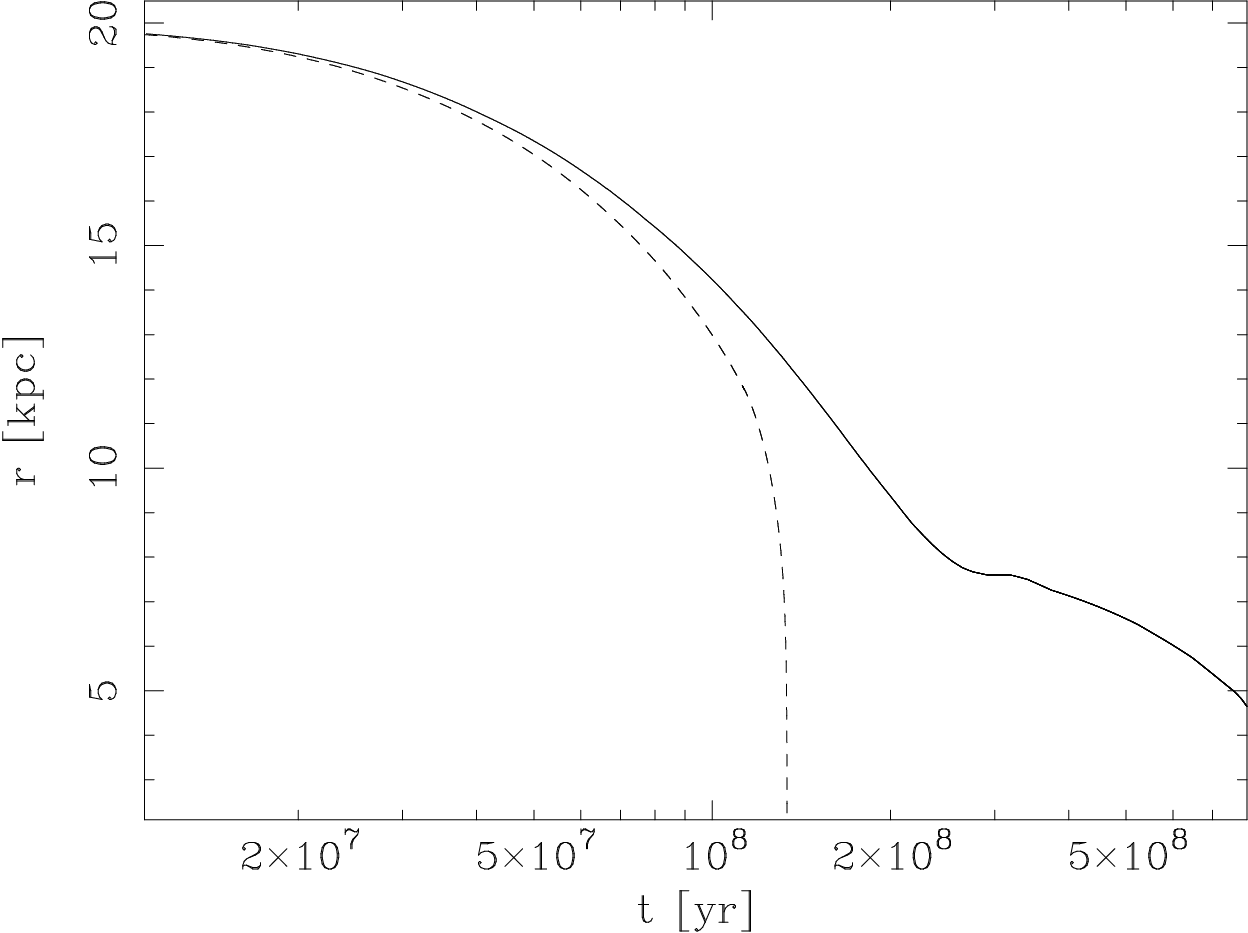}
\includegraphics[width =85mm, angle = -90]{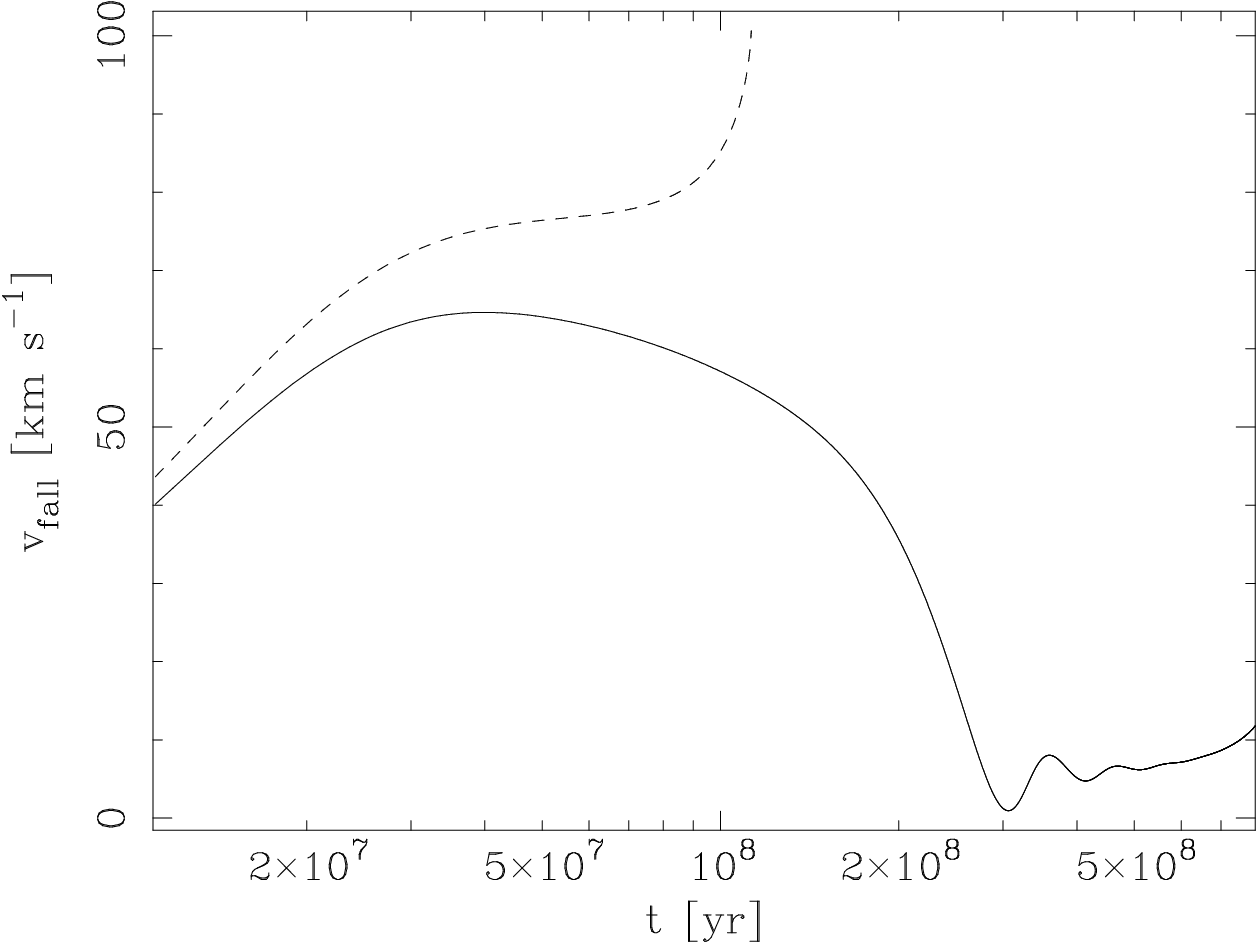}
\caption{\label{f:rv}
The distance from the cluster center (upper panel) and the fall
velocity (lower  panel)
of two blobs with initial over-densities $\delta_0=2.0$
(solid line) and $\delta_0=2.5$ (dashed line). Both the blobs have the
initial radius $a_0=100$~pc, and have been  released from rest
at the distance $r_0 = 20$~kpc from the cluster center.
}
\end{figure}

\begin{figure}
\centering
\includegraphics[width =85mm, angle = -90]{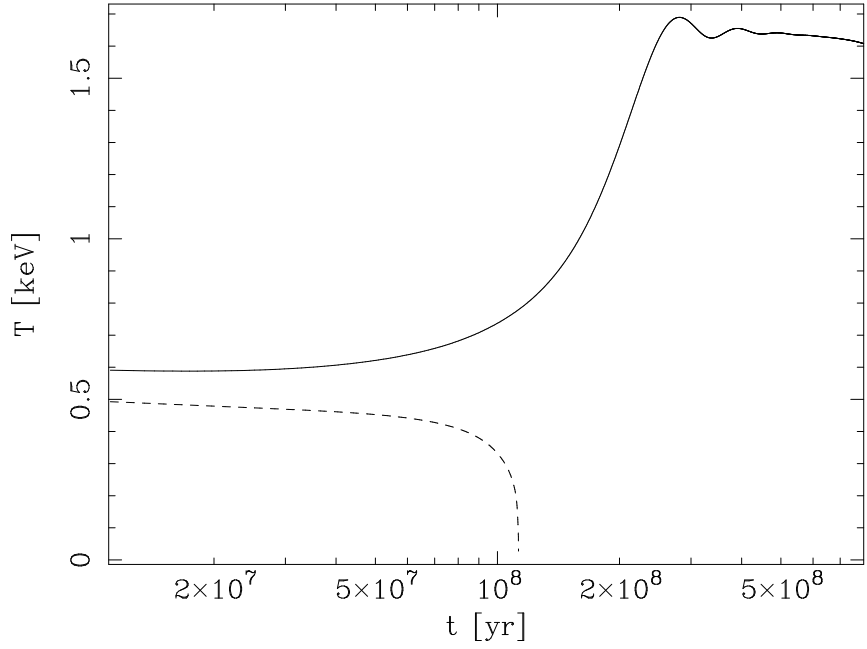}
\caption{\label{f:temp}
The temperature evolution  of two blobs with the same characteristics
as in Figure~\ref{f:rv}.  Also the line styles are the same as in
Figure~\ref{f:rv}.
}
\end{figure}

\begin{figure}
\centering
\includegraphics[width =85mm, angle = -90]{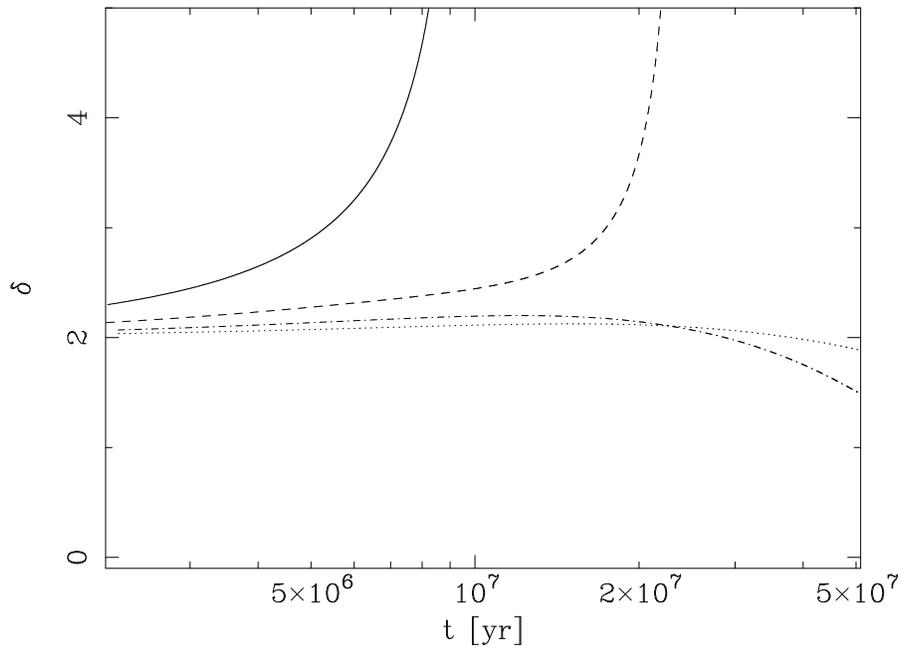}
\caption{\label{f:radii}
The evolution of the over-density as a function of the release radius:
$r=2$~kpc (solid line),
$r=5$~kpc (dashed line),
$r=10$~kpc (dash-dotted line), and
$r=20$~kpc (dotted line).
In all  cases the blob has an initial over-density $\delta_0=2$, and an initial
radius $a_0=100$~pc.
}
\end{figure}

\begin{figure}
\centering
\includegraphics[width =85mm, angle = -90]{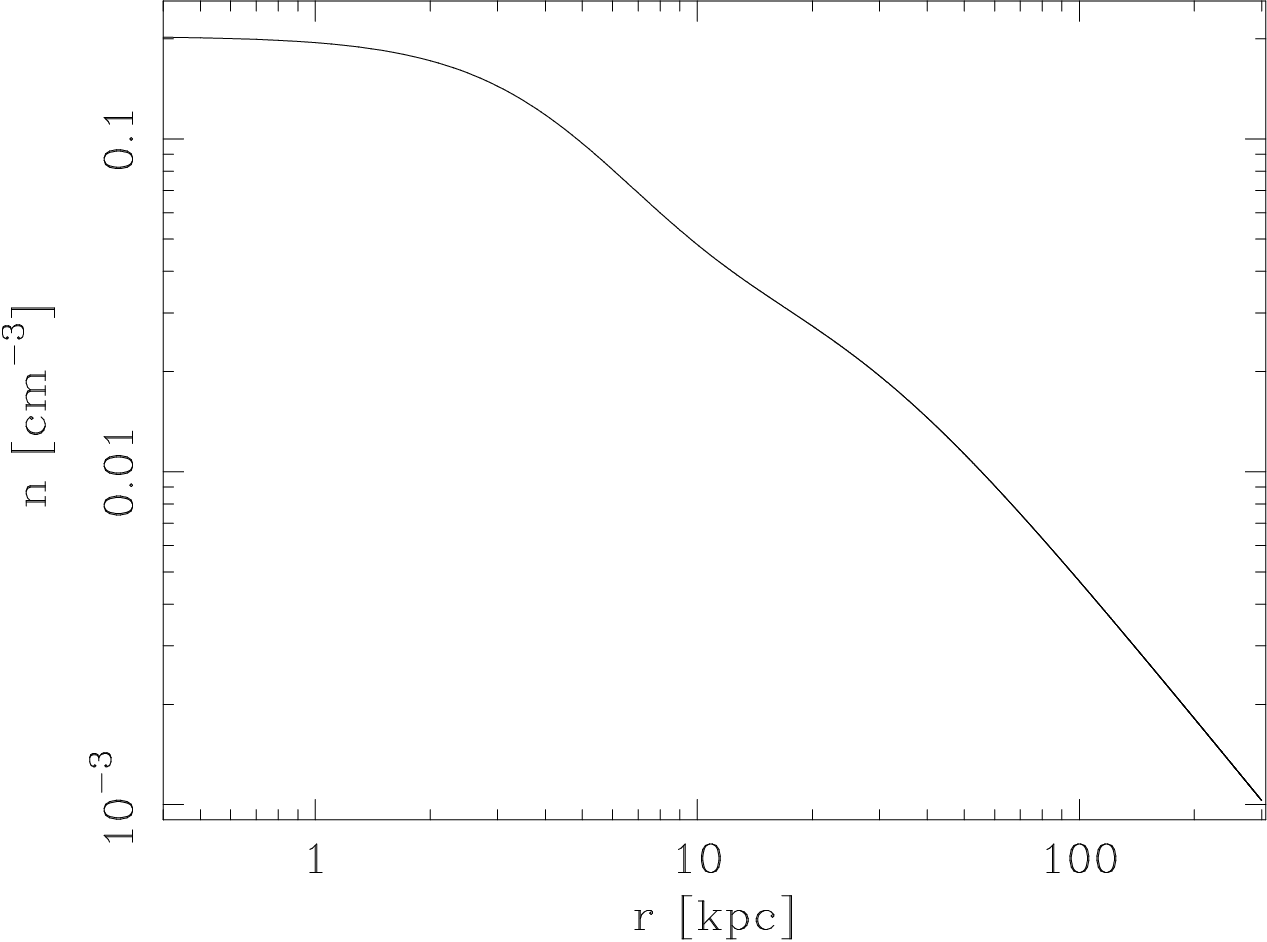}
\includegraphics[width =85mm, angle = -90]{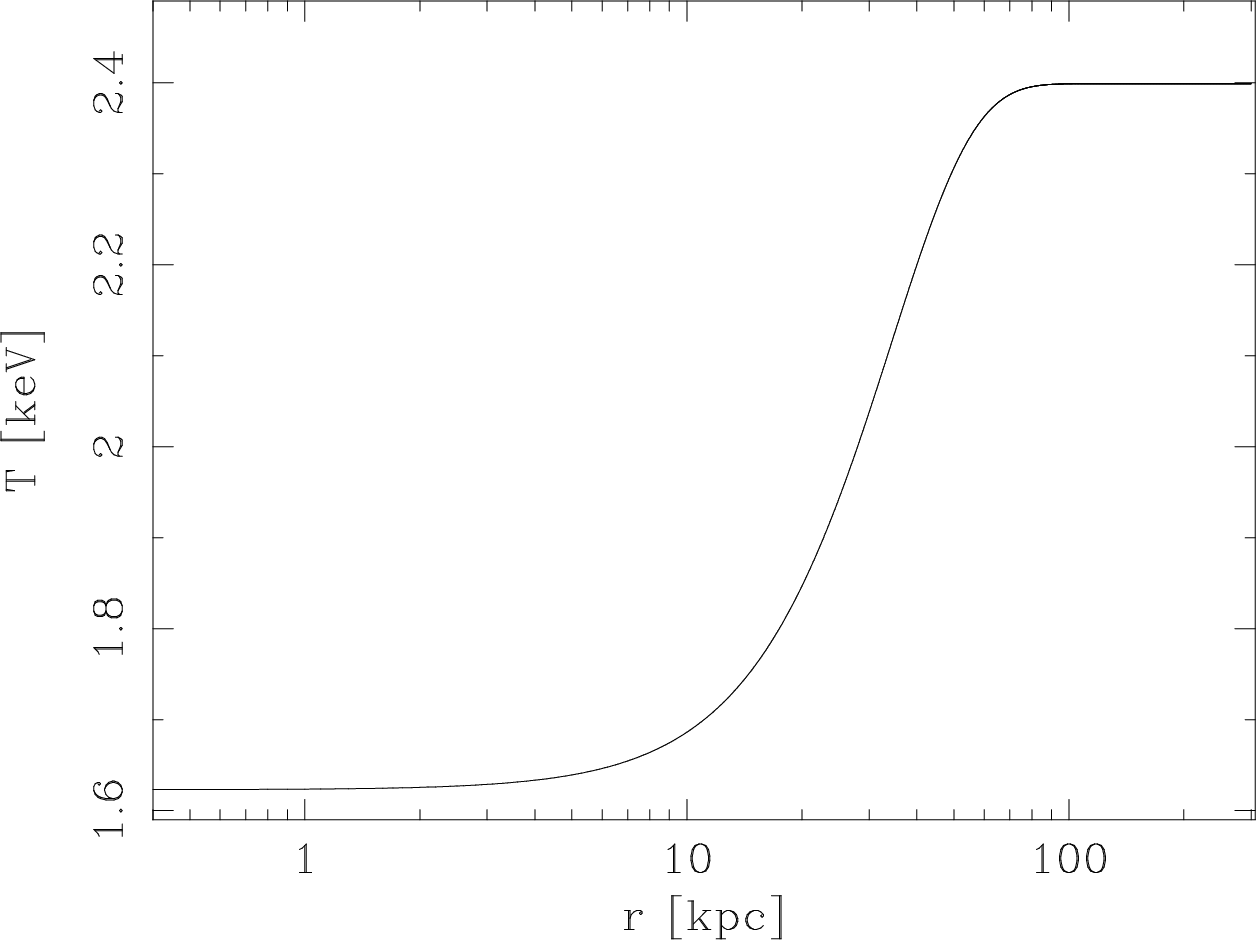}
\caption{\label{f:gas}
The total number particle density (upper panel) and the temperature profile
(lower panel) of the ambient gas in M87. These profiles have been
 been borrowed from \citet{Ghi04}.
}
\end{figure}

\begin{figure}
\centering
\includegraphics[width =85mm, angle = -90]{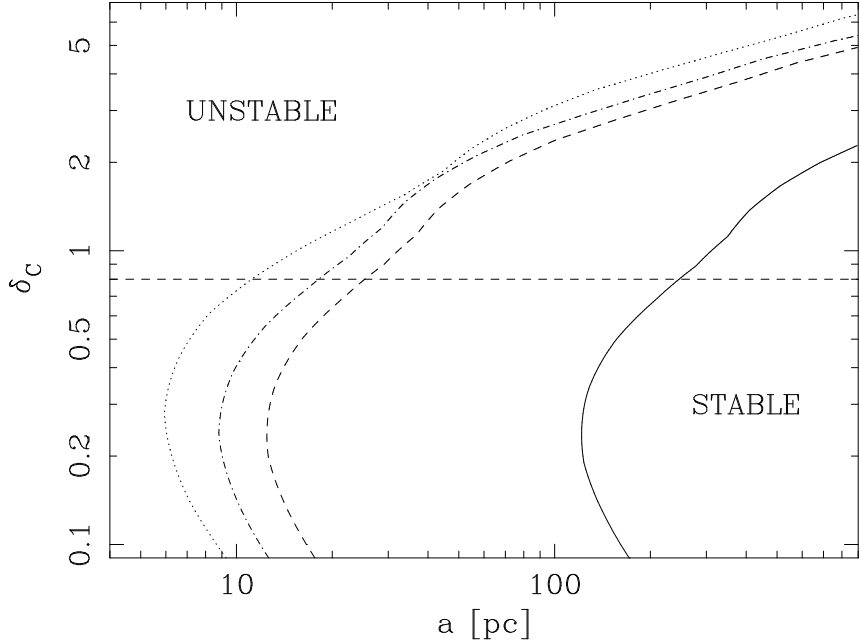}
\caption{\label{f:deltacrit}
The critical density $\delta_C$ as a function of the blob size $a$
for different initial positions of the blob:
$r=2$~kpc (solid line),
$r=5$~kpc (dashed line),
$r=10$~kpc (dash-dotted line), and
$r=40$~kpc (dotted line).
The blobs on the left part of the plot are unstable, and  evolve to larger
values of $\delta\rho/\rho_a$. The blobs on the right part are stable, and
evolve to smaller over-densities.
The blobs with sizes in the range $a \approx 10-100$~pc are unstable
provided that their over-densities are in the moderately non-linear regime
$\delta\rho/\rho_a\approx 0.8-3$, or the that they have been released from
very small radii. For consistence with our neglect of the
cooling of the ambient gas, we should consider only the part of the curves
above the horizontal dashed line, corresponding to a blob cooling
time of about $30\%$~of the ambient  gas. Blobs with over-densities below
this (rather arbitrary) limit have cooling times closer to the ambient gas.
The $30\%$ confidence limit  has been calculated by comparing the
cooling time $t_a\propto T_a/n_a\Lambda(T_a)$ for the ambient gas with  the
corresponding expression for the blob cooling time $t_b$.
We assume pressure equilibrium between the ambient and the blob, so
$T_a = (1+\delta)\: T_b$, where $T_b$ is the blob temperature.
By approximating  the cooling function with  a power law
$\Lambda\propto T^\alpha$, we obtain $t_a/t_b = (1+\delta)^{2-\alpha}$.
Our approximation $t_b \ll t_a$ requires $\delta\gg 1$. If we
demand $t_b < 30\%\: t_a$ and take  $\alpha \approx 0$
in the temperature range
considered here, we obtain the plotted confidence limit $\delta > 0.8$.
}
\end{figure}

Most previous papers (see \S~\ref{s:intro}), assume that the central 
black hole
accretes mass only from the ICM in its immediate
neighborhood, basically  via a Bondi-like accretion flow.
In these models the feedback occurs as the ICM cools to a temperature
of about $1 \keV$, and the ICM does not need to cool to low temperatures.
We term these {\it hot-feedback models}.
We examined three papers  based on hot-feedback
heating  \citep[][]{Omm04, Rus02, Hoe04}.
We argued that the models worked out in these papers do not fit the general
temperature
profiles of CFs (Figure~\ref{f:tprof}), and/or require fine tuning.
We further argued that more generally, in the Bondi-type accretion
flow of hot gas, the accretion rate is determined mainly by the
conditions very close to the central black hole, and that this may
result in unstable cooling of the regions further out.

We therefore proposed (\S~\ref{s:scenario}) that the feedback occurs with
the entire cool inner region, $r \lesssim 5-30 \kpc$, in
what we term a {\it cold-feedback model}.
In the proposed scenario non-linear over-dense
($\delta \rho /\rho_a \gtrsim 2$, or $\rho/\rho_a \gtrsim 3$)
blobs of gas cool fast and  are removed from
the ICM before  the next major AGN heating event in their region.
It is important to note that an AGN burst can take place and
heat other regions, since the jets and/or bubbles may
expand in other directions as well.
The typical interval between such heating events at a specific region
is~$\approx 10^8 \yr$.
Some of these blobs cool and sink toward the central black hole,
while others may form stars and cold molecular clouds.

Four conditions should be met in the inner region participating
in the feedback heating.
\begin{enumerate}
\item In order for the blob not to reach a
point where its density equals the ambient density as it sinks,
the ICM density profile should not be too steep. This implies that
the relevant dense blobs form in the cluster core, where the density
profile is shallow. In the quantitative example used here
for M87 this is the region $r \lesssim 5 \kpc$, while in more
typical clusters it is larger, e.g., $r \lesssim 30 \kpc$ in
A2052. We note that the lower segment of magnetic flux loops can
be prevented from reaching the stabilizing point by the upward
force of the magnetic tension inside the loop \citep[][]{Sok04a}.
Therefore, some perturbations can be formed at large distances,
where density profile is steep, and still cool to low temperature
and feed the central black hole.
\item Non-linear perturbations  are required. These presumably formed mainly
by previous AGN activity, e.g.  jets and radio lobes.
\item The cooling rate  of these non-linear perturbations
is short relative to few times the typical interval  between
successive AGN outbursts.
\item
The blobs must not be evaporated by thermal conduction
before they are delivered to the AGN. This requires a strong
suppression of thermal conduction, which may be done by the 
magnetic fields observed in the radio lobes in several cooling
flow clusters.

\end{enumerate}
The first and the third condition, which are not completely
independent of each other,
require that the initial ICM cools by a factor of a few before the
feedback starts operating, and the second condition requires that the
inner region must be disturbed.

Finally, in \S~\ref{s:nnlin} we have calculated the falling time 
and cooling time of dense blobs. The results have then been  applied to the 
cooling-flow  cluster M87.

The cold-feedback model has the following implications and predictions
(for more details and references see \S~\ref{s:intro}).
\begin{enumerate}
\item
In the cold-feedback models the optical filaments
observed in many CF-clusters and the cooler molecular
gas detected via CO observations come from cooling ICM
(with some amount possibly from stripping from galaxies).
\item
In cold feedback models, some X-ray emission from gas at
temperatures $\lesssim 10^7$~K is predicted to exist,
much more than in many other AGN heating models.
but at a level more than an order of magnitude below that
in old versions of the CF model, but compatible with the
moderate CF model.
We stress that in the cold-feedback heating, cooling flows
{\it  do exist}. 
Such gas cooling to below X-ray emitting temperatures was
found recently in two CF clusters 
(Abell~2597: \citealt{Mor05}; Abell~2029: \citealt{Cla04}).

\item
The feeding of the central black hole with cold gas in the
cold feedback models makes the process similar in some
aspects to that of AGN in spiral galaxies.
Therefore, the outflow can be similar \citep[][]{Sok05}.

\item
It is possible that in the cold feedback model a
substantial fraction of gas that cooled to low
temperatures and was accreted to the accretion disc around
the central black hole, is injected back to the ICM at
non-relativistic velocities \citep[][]{Sok05}.

\end{enumerate}


\acknowledgments
This work is supported by a grant of the Israeli
Science Foundation.  FP acknowledges a Fine Fellowship. 
FP was supported also by grant No.~2002111
from the United States-Israel Binational Foundation (BSF), Jerusalem, Israel.
We  thank an anonymous referee for helpful comments which
improved the presentation of our results.


\clearpage



\newpage

\end{document}